\documentclass[12pt]{article}
\usepackage{amsmath}
\usepackage{amssymb}
\usepackage{graphics}

\def\pa{\partial}
\def\del{\partial}
\def\MP{{\mathrm{MP}}}

\usepackage[dvips]{color}
\usepackage[dvips]{graphicx}
\usepackage[normalem]{ulem}
\usepackage[dvips]{color}

\begin{document}
\begin{titlepage}
\begin{flushright}
TIT/HEP-616\\
October 2011
\end{flushright}
\vspace{0.5cm}
\begin{center}
{\Large \bf
Deformed BPS Monopole in $\Omega$-background}
\lineskip .75em
\vskip1.0cm
{\large Katsushi Ito${}^{*}$, Satoshi Kamoshita${}^{*}$ and Shin Sasaki${}^{\dagger}$ }
\vskip 2.5em
${}^{*}$ {\normalsize\it Department of Physics\\
Tokyo Institute of Technology\\
Tokyo, 152-8551, Japan} \vskip 1.5em
${}^{\dagger}$ {\normalsize\it Department of Physics\\
Kitasato University\\
Sagamihara, 228-8555, Japan} \vskip 1.5em
\end{center}
\begin{abstract}
We study the BPS condition in the
$\Omega$-deformed $\mathcal{N}=2$ super Yang-Mills theory
when one of the $\epsilon$-parameters of the background is zero.
We obtain the deformed BPS equation for dyons
and the formula for their central charge.
In particular, we find that the deformed BPS monopole
equation has axially-symmetric solution
and is equivalent to the Ernst equation.
The monopole charge is shown to be undeformed.
We construct one-monopole solution explicitly
and examine its profile.
\end{abstract}
\end{titlepage}

\baselineskip=0.7cm
%\tableofcontents

The $\Omega$-background deformation of $\mathcal{N}=2$ supersymmetric
gauge theories \cite{Moore:1997dj,Nekrasov:2002qd,Losev:2003py}
has been attracted much attentions.
The $\Omega$-background is characterized by two
complex parameters $\epsilon_1$ and $\epsilon_2$
associated with the $U(1)^2$ action on four-dimensional spacetime
${\bf R}^4= {\bf R}^2 \times {\bf R}^2$.
This background breaks the super-Poincar\'e invariance in general. 
By introducing the R-symmetry gauge field Wilson line,
the deformed theory has one equivariantly nilpotent supercharge.
The instanton partition function can be evaluated
via the localization theorem by using the supercharge.
One can reproduce the Seiberg-Witten (SW) prepotential \cite{SW}
from the instanton partition function by taking the limit
$\epsilon_1,\epsilon_2 \rightarrow 0$~\cite{Nekrasov:2002qd}.

The theory has extended supersymmetries in two dimensions
when one of the $\epsilon$-parameters is set to be zero.
For $\epsilon_2=0$, the instanton partition function
gives the prepotential deformed by $\epsilon_1$.
The deformed prepotential is related to the
Yang-Yang functional in a quantum integrable system,
where $\epsilon_1$ plays a role of the Planck constant
\cite{Nekrasov:2009rc, Nekrasov:2010ka, Maruyoshi:2010iu,
	Orlando:2010uu}.
Recently, it has been pointed out that
the prepotential can be also evaluated by
the period integral of the deformed SW differential
obtained from the quantization of the SW curve
\cite{Poghossian:2010pn}.
The periods are identified with the exact
Bohr-Sommerfeld integral of the 1d sine-Gordon model
\cite{Mironov:2009uv, Mironov:2009dv, Mironov:2009ib,Popolitov:2010bz}.

The deformation of the prepotential should be also derived from the 
microscopic calculation of the deformed super Yang-Mills theory.
In the SW theory, the period integrals
of the SW differential are the central charges of the 
supersymmetry algebra, which is related to the
masses of the BPS states \cite{Witten:1978mh}.
In this paper we will investigate the BPS states in the $\Omega$-deformed
theory in order to study the deformed
prepotential from the field theoretical point of view.
In the previous paper \cite{Ito:2011ta},
we have shown that the theory with $\epsilon_1=0$ or $\epsilon_2=0$
has the deformed
$\mathcal{N}=(2,2)$ or $(2,1)$
supersymmetry\footnote{We denote $\mathcal{N}=(p,q)$ by supersymmetry
with $p$ chiral and $q$ anti-chiral supercharges.},
which depends on the choice of the Wilson lines. We also derived the formula for its
central charge and the BPS monopole
equation which preserves part of the
supersymmetries.
In this paper we will consider more general BPS
configurations, which give the BPS equation for dyon.
We will obtain the central charge for the BPS dyon,
which includes the monopole charge as an example.
We will then study the BPS monopole solution
and evaluate its charge.
For the construction of the solution,
it is useful to refer to
axially-symmetric monopole solutions in the Yang-Mills-Higgs model
\cite{Prasad:1975kr, Bog}
based on the
Manton's ansatz \cite{Ma}.
Forg\'acs et al. \cite{Forgacs:1980yv} showed that
the monopole equation with this ansatz reduces to the Ernst
equation \cite{Ernst:1967wx}. It can be solved by the inverse scattering method \cite{Belinsky}
and the B\"acklund transformation \cite{Harrison,Neugebauer}.
They found that one obtains
multi-monopole solutions by applying multi-consecutive
B\"acklund transformations to a simple vacuum solution
\cite{Forgacs:1981jt}. 
In this paper we will show that the
deformed BPS monopole equation
is equivalent to the same Ernst equation.
We will also solve the BPS equation associated with one-monopole
perturbatively in $\epsilon_1$.

We consider $U(N)$ $\mathcal{N}=2$ super Yang-Mills theory in
the $\Omega$-background.
The theory contains a gauge field $A_{m}$ ($m=1,2,3,4$), 
Weyl fermions $\Lambda^{I}$, $\bar{\Lambda}^{I}$ ($I=1,2$), 
and complex scalars $\varphi$, $\bar{\varphi}$. They belong to 
the adjoint representation of $U(N)$ gauge group.
Here $I$ denotes an $SU(2)_I$ R-symmetry indices.
We also introduce R-symmetry gauge field Wilson lines ${\cal
A}^{I}{}_J$ and $\bar{\cal A}^I{}_J$.
The Lagrangian is given by \cite{Ito:2010vx}
\begin{align}
\mathcal{L}^{}_{\Omega}
&=
\frac{1}{g^2\kappa}\textrm{Tr}\biggl[
\frac{1}{4}F_{mn}F^{mn}
+(D_{m}\varphi-F_{mn}\Omega^{n})
(D^{m}\bar{\varphi}-F^{mp}\bar{\Omega}_{p})
\notag\\
&\qquad\qquad{}
+\Lambda^{I}\sigma^{m}D_{m}\bar{\Lambda}_{I}
-\frac{i}{\sqrt{2}}\Lambda^{I}[\bar{\varphi},\Lambda_{I}]
+\frac{i}{\sqrt{2}}\bar{\Lambda}_{I}[\varphi,\bar{\Lambda}^{I}]
\notag\\
&\qquad\qquad{}
+\frac{1}{\sqrt{2}}\bar{\Omega}^{m}\Lambda^{I}
D_{m}\Lambda_{I}
-\frac{1}{2\sqrt{2}}\bar{\Omega}^{mn}\Lambda^{I}
\sigma_{mn}\Lambda_{I}
\notag\\
&\qquad\qquad{}
-\frac{1}{\sqrt{2}}\Omega^{m}\bar{\Lambda}_{I}
D_{m}\bar{\Lambda}^{I}
+\frac{1}{2\sqrt{2}}\Omega^{mn}\bar{\Lambda}_{I}
\bar{\sigma}_{mn}\bar{\Lambda}^{I}
\notag\\
&\qquad\qquad{}
+\frac{1}{2}\Bigl([\varphi,\bar{\varphi}]
+i\Omega^{m}D_{m}\bar{\varphi}-i\bar{\Omega}^{m}D_{m}\varphi
+i\bar{\Omega}^{m}\Omega^{n}F_{mn}
\Bigr)^{2}
\notag\\
&\qquad\qquad{}
-\frac{1}{\sqrt{2}}\bar{\mathcal{A}}^{J}{}_{I}
\Lambda^{I}\Lambda_{J}
-\frac{1}{\sqrt{2}}\mathcal{A}^{J}{}_{I}
\bar{\Lambda}^{I}\bar{\Lambda}_{J}
\biggr],
\label{omega_lag}
\end{align}
where $F_{mn}=\partial_{m}A_{n}-\partial_{n}A_{m}+i[A_{m},A_{n}]$ 
is the gauge field strength, 
$D_{m}=\partial_{m}+i[A_{m},\ast]$ 
is the gauge covariant derivative,
$g$ is the gauge coupling constant
and the constant $\kappa$ normalizes
basis of $U(N)$.
We will consider the Lagrangian in Euclidean spacetime. 
We then define the Dirac matrices 
$\sigma_{m}=(i\tau^1,i\tau^2,i\tau^3,1)$ and
$\bar{\sigma}_{m}=(-i\tau^1,-i\tau^2,-i\tau^3,1)$,
where $\tau^{c}$ ($c=1,2,3$) are the Pauli matrices. 
$\sigma_{mn}$ and $\bar{\sigma}_{mn}$ are the Lorentz generators.
%The Lorentz generators $\sigma^{mn}$ and $\bar{\sigma}^{mn}$ are defined
%by
%\begin{gather}
%(\sigma^{mn})_{\alpha}{}^{\beta}=
%\frac{1}{4}\bigl(
%\sigma^{m}_{\alpha\dot{\alpha}}\bar{\sigma}^{n\dot{\alpha}\beta}
%-\sigma^{n}_{\alpha\dot{\alpha}}\bar{\sigma}^{m\dot{\alpha}\beta}
%\bigr),
%\quad 
%(\bar{\sigma}^{mn})^{\dot{\alpha}}{}_{\dot{\beta}}=
%\frac{1}{4}\bigl(
%\bar{\sigma}^{m\dot{\alpha}\alpha}\sigma^{n}_{\alpha\dot{\beta}}
%-\bar{\sigma}^{n\dot{\alpha}\alpha}\sigma^{m}_{\alpha\dot{\beta}}
%\bigr).
%\end{gather}
We set the vacuum theta-angle to zero for simplicity.  
The $\Omega$-background is parametrized by $\Omega_{mn}$
and $\bar{\Omega}_{mn}$ which take the form
\begin{align}
\begin{aligned}
\Omega^{mn}=\frac{1}{2\sqrt{2}}
\begin{pmatrix}
0 & i\epsilon_1 & 0 & 0  \\
-i\epsilon_1 & 0 & 0 & 0 \\
0 & 0 & 0 & -i\epsilon_2 \\
0 & 0 & i\epsilon_2 & 0  \\
\end{pmatrix}, &\quad
\bar{\Omega}^{mn}=\frac{1}{2\sqrt{2}}
\begin{pmatrix}
0 & -i\bar{\epsilon}_1 & 0 & 0  \\
i\bar{\epsilon}_1 & 0 & 0 & 0 \\
0 & 0 & 0 & i\bar{\epsilon}_2 \\
0 & 0 & -i\bar{\epsilon}_2 & 0  \\
\end{pmatrix}.
\end{aligned}
\label{omegab}
\end{align}
The vector fields $\Omega^m$ and $\bar{\Omega}^m$ are defined by 
$\Omega^m = \Omega^{mn}x_n$ and
$\bar{\Omega}^m = \bar{\Omega}^{mn}x_n$.
They generate $U(1)\times U(1)$ actions on ${\bf R}^4$.
Two parameters $\epsilon_1, \epsilon_2$ break the
super-Poincar\'e invariance in general.
However, when one of the $\epsilon$-parameters $\epsilon_1$ or
$\epsilon_2$ becomes zero, the super-Poincar\'e
invariance over the $(x^1,x^2)$-plane
or the $(x^3,x^4)$-plane
recovers respectively~\cite{Nekrasov:2009rc}.

Supersymmetries in the limit $\epsilon_1$ or
$\epsilon_2 \to 0$ have been studied
in \cite{Ito:2011ta}. They depend on the choice
of the Wilson lines.
To see them, it is convenient to
introduce the topological twist
\cite{Witten:1988ze}.
We identify the $SU(2)_I$ R-symmetry index with the $SU(2)_R$ spinor
index in the Lorentz group $SO(4) = SU(2)_L \times SU(2)_R$. 
The twisted supercharges are defined as 
\begin{eqnarray}
Q_m = (\bar{\sigma}_m)^{I \alpha} Q_{\alpha I}, \quad 
\bar{Q} = \delta^{\dot{\alpha}} {}_I \bar{Q}^I {}_{\dot{\alpha}}, \quad 
\bar{Q}_{mn} = - (\bar{\sigma}_{mn})^{\dot{\alpha}} {}_I \bar{Q}^I {}_{\dot{\alpha}},
\end{eqnarray}
where $Q_{\alpha I}$ and $\bar{Q}^I {}_{\dot{\alpha}}$ are supercharges
associated with the $\mathcal{N} = (4,4)$ supersymmetry in the
undeformed theory.
The $\mathcal{N}=(4,4)$ supersymmetry algebra reads
\begin{eqnarray}
\begin{aligned}
& \{Q_m, \bar{Q} \} = 4 P_m, \ \{Q_m, \bar{Q}_{pq} \} = 2 (\delta^{nq}
\delta^{mp} - \delta^{np} \delta^{mq} - \varepsilon^{mnpq}) P_n, \\
& \{\bar{Q}, \bar{Q}\} = 4 \sqrt{2} \bar{Z}, \ \{Q_m, Q_n\} = - 4
\sqrt{2} \delta_{mn} Z, \\
& \{\bar{Q}_{mn}, \bar{Q}\} = 0, \ \{\bar{Q}_{mn}, \bar{Q}_{pq} \} =
\sqrt{2} (\delta^{mp} \delta^{nq} - \delta^{mq} \delta^{np} -
\varepsilon^{mnpq}) \bar{Z},
\end{aligned}
\end{eqnarray}
where $P^m$ is the four-momentum and $Z, \bar{Z}$ are the central
charges. When the Wilson line is
\begin{eqnarray}
\mathcal{A}^I {}_J = - \frac{1}{2} \Omega_{mn} (\bar{\sigma}^{mn})^I
{}_J, \quad 
\bar{\mathcal{A}}^I {}_J = - \frac{1}{2} \bar{\Omega}_{mn}
(\bar{\sigma}^{mn})^I {}_J, 
\end{eqnarray}
the theory has $\mathcal{N} = (2,1)$ supersymmetry which is
generated by
$Q^1, Q^2, \bar{Q}$ in the case of $\epsilon_1=0$ and by 
$Q^3, Q^4, \bar{Q}$ in the case of $\epsilon_2=0$.
On the other hand, when the Wilson line is
\begin{align}
\mathcal{A}^{I}{}_{J}
=-\frac{1}{2}\bar{\Omega}_{mn}(\bar{\sigma}^{mn})^{I}{}_{J},\quad 
\bar{\mathcal{A}}^{I}{}_{J}
=-\frac{1}{2}\Omega_{mn}(\bar{\sigma}^{mn})^{I}{}_{J},
\end{align}
the theory has $\mathcal{N} = (2,2)$ supersymmetry which is generated by
$Q^3, Q^4, \bar{Q}^{13}, \bar{Q}^{14}$ in the case of $\epsilon_1=0$ and by
$Q^1, Q^2, \bar{Q}^{13}, \bar{Q}^{14}$ in the case of $\epsilon_2=0$.
We note that the $\mathcal{N}=(2,1)$ and $\mathcal{N}=(2,2)$
transformations for the fields
are deformed by the remaining $\epsilon$-parameter.

We now examine
the BPS equation in the $\Omega$-background
for the dyonic state from the energy
bound. To find the bound, we go back to
the Minkowski spacetime and use the phase transformation of 
$\varphi$ and $\Omega$ to set those to
$\varphi = - \bar{\varphi}, \Omega =- \bar{\Omega}$
and define $\hat{\Omega} = \sqrt{2} i \Omega$,
$\phi =\sqrt{2} i \varphi$ such that $\phi$ and $\Omega$ are real values.
Performing the Bogomol'nyi completion,
the energy becomes
\begin{eqnarray}
E &=& 
\int \! d^3 x \ \frac{1}{\kappa} \mathrm{Tr} 
\left[
\frac{1}{2} 
\left\{
E_i \pm (D_i \phi +  \hat{\Omega}^j F_{ji}) \sin \theta
\right\}^2
+ 
\frac{1}{2}
\left\{
B_i \pm (D_i \phi +  \hat{\Omega}^j F_{ji}) \cos \theta
\right\}^2
\right. \nonumber \\
& & \qquad \qquad \qquad 
\left.
+ \frac{1}{2} (D_0 \phi +  \hat{\Omega}^j F_{j0})^2
\right]
\nonumber \\
& & \mp \frac{1}{\kappa} \int \! d^3 x 
\ \mathrm{Tr}
\left[
B^i D_i \phi
\right] \cos \theta 
\mp \frac{1}{\kappa}
\int \! d^3 x 
\mathrm{Tr}
\left[
E^i (D_i \phi +  \hat{\Omega}^j F_{ji}) \sin \theta
\right],
\label{energy_density}
\end{eqnarray}
where $E_i = F_{i0}$ $(i=1,2,3)$ is the electric field, $B_i = \frac{1}{2} \epsilon_{ijk}
F^{jk}$ is the magnetic field and $\theta$ is an arbitrary parameter. 
Here $x_0 = - i x_4$ and other vectors
are defined in a similar way.

The energy bound is given by the last two terms in
\eqref{energy_density}.
The energy is saturated provided that the
following BPS conditions are satisfied:
\begin{eqnarray}
\begin{aligned}
& E_i \pm (D_i \phi + \hat{\Omega}^j F_{ji} ) \sin \theta = 0, 
\\
& B_i \pm (D_i \phi + \hat{\Omega}^j F_{ji} ) \cos \theta = 0, 
\\
& D_0 \phi +  \hat{\Omega}^j F_{j0} = 0. 
\end{aligned}
\label{BPS_eq}
\end{eqnarray}
The last equation in (\ref{BPS_eq}) and the equations of motion
for the electric field imply Gauss' law $D_i E_i = 0$.
Using the Bianchi identity $D_i B_i = 0$ and Gauss' law, 
the energy bound is rewritten as 
\begin{align}
E= \mp \{ Q_\mathrm{m} \cos \theta + (Q_\mathrm{e} + \delta Q_\mathrm{e}) \sin \theta \},
\label{dyon_mass}
\end{align}
where $Q_\mathrm{m}, Q_\mathrm{e}$ are the undeformed magnetic and electric charges defined by 
\begin{equation}
Q_\mathrm{m} = \frac{1}{\kappa} \int \! d^3 x \ \partial_i \mathrm{Tr}
[B_i \phi], \quad 
Q_\mathrm{e} = \frac{1}{\kappa} \int \! d^3 x \ \partial_i \mathrm{Tr} 
[E_i \phi],
\label{magch}
\end{equation}
while $\delta Q_\mathrm{e}$ denotes the correction to the electric charge:
\begin{eqnarray}
\label{deltaQ}
\delta Q_\mathrm{e} = \frac{1}{\kappa} \int \! d^3 x \ \mathrm{Tr} 
[ \hat{\Omega}^j F_{ji} E^i].
\end{eqnarray}
The energy bound is minimized when the parameter $\theta$ 
satisfies the following condition 
\begin{eqnarray}
\sin \theta = \frac{Q_\mathrm{m}}{\sqrt{Q_\mathrm{m}^2 + Q_\mathrm{e}^{\prime 2}}},
\label{theta}
\end{eqnarray}
where we have defined the deformed electric charge
$Q_\mathrm{e}^{\prime} = Q_\mathrm{e} + \delta Q_\mathrm{e}$.
When this condition is satisfied, the energy is given by the mass of the
BPS state, namely, the dyon mass, 
\begin{eqnarray}
M_{\mathrm{dyon}} = \sqrt{Q_\mathrm{m}^2 + Q^{\prime 2}_\mathrm{e}}.
\end{eqnarray}

To see the relation between the energy bound and the central charge in
the supersymmetry algebra, we evaluate the deformed central charge by calculating
the anti-commutation relation of the Noether charges associated with the
deformed supersymmetry transformation in Euclidean space \cite{Ito:2011ta}.
We have found that
\begin{align}
%Z = \int \! d^3 x \ \frac{1}{\sqrt{2}}
Z = \int \! d^3 x \ \frac{1}{\sqrt{2}\kappa}
\mathrm{Tr}
\left[
i D_i \phi B_i - D_i \phi E_i - \hat{\Omega}^j F_{ji} E_i
\right], 
\label{central_charge}
\end{align}
where we have used the BPS conditions 
\eqref{BPS_eq} and the parameter (\ref{theta}).
The mass of the BPS state defined by
the deformed BPS equations \eqref{BPS_eq} 
is given by the deformed central charge:
\begin{eqnarray}
M_{\mathrm{dyon}} = \sqrt{Q_\mathrm{m}^2 + Q_\mathrm{e}^{\prime 2}} = \sqrt{2} |Z|.
\end{eqnarray}
We note that in the BPS monopole state $B_i \not= 0, E_i = 0$, 
the expression of central charge is not deformed by the
$\Omega$-background.
However, this can be deformed through the deformed
solution of the monopole equation.

The dyon BPS state preserves parts of deformed supersymmetries.
Substituting the BPS conditions 
\eqref{BPS_eq} into the deformed 
supersymmetry transformation of
fermions in \cite{Ito:2011ta}, we get the
following condition
\begin{eqnarray}
e^{- i \theta} \xi_m (\sigma^m)_{\alpha I} 
\mp i (\sigma^{mn})^{\dot{\alpha}} {}_I \bar{\xi}_{mn} \mp i
\delta^{\dot{\alpha}} {}_I \bar{\xi} = 0,
\end{eqnarray}
where $\xi_m, \bar{\xi}_{mn}, \bar{\xi}$ are transformation parameters
associated with the supercharges $Q_m, \bar{Q}_{mn}$, and $\bar{Q}$.
This condition is the same as the monopole case
provided that $\xi_m$ is replaced by $e^{-i \theta} \xi_m$.

We consider the case $\epsilon_2=0$ where the BPS equation
preserves at least one supercharge. Our purpose is to evaluate
the $\epsilon_1$-correction to the central charge.
In this case the central charge \eqref{central_charge} becomes
\begin{align}
% Z = \int \! d^3 x \ \frac{1}{\sqrt{2}}
Z = \int \! d^3 x \ \frac{1}{\sqrt{2}\kappa}
\mathrm{Tr}
\left[
i ({\boldsymbol D} \phi) \cdot {\boldsymbol B} - ({\boldsymbol D} \phi)
\cdot {\boldsymbol E} - \epsilon  ({\boldsymbol x}
\times ( {\boldsymbol E} \times {\boldsymbol B}) )_3
\right],
\label{eq:central2}
\end{align}
where we have defined $\epsilon= -\mathrm{Re}(\epsilon_1)/2$ and
introduced the three-vectors ${\boldsymbol x} =(x^1,x^2,x^3)$ etc.
In order to evaluate it,
it is necessary to solve the dyon equation and substitute the
solution into the central charge.
In this paper, we consider the deformed
BPS monopole equation which is
the simplest example.
The deformed BPS monopole equation is obtained by
setting $\theta=0,\ A_0=0,\ \del_0\phi=\del_0 A_i=0$
in the equations \eqref{BPS_eq}:
\begin{align}
B_i  \pm \left( D_i \phi + \hat{\Omega}^j F_{ji} \right)=0.
\label{monopole}
\end{align}
Hereafter we consider the $SU(2)$ gauge group for simplicity.
For $\epsilon_2=0$, we see that the deformed BPS equation
has the axial symmetry around the $x^3$-axis.
Hence we use Manton's ansatz \cite{Ma} for the fields:
\begin{align}
\begin{aligned}
A^a_i&=\left\{ \eta_1\hat{\rho}^a+\left( \eta_2 + \frac{1}{g\rho} \right) \hat{z}^a \right\}\hat{\varphi}^i
+W_1\hat{\rho}^i\hat{\varphi}^a+W_2\hat{z}^i \hat{\varphi}^a,\\
\phi^a&=\phi_1\hat{\rho}^{a}+\phi_2\hat{z}^a,
\end{aligned}
\label{mansatz}
\end{align}
where $(\rho,\varphi,z)$ are the cylindrical coordinates
for the spatial direction $(x^1,x^2,x^3)$,
$\hat{\rho}=(\cos\varphi,\sin\varphi,0),
\ \hat{\varphi}=(-\sin\varphi,\cos\varphi,0),
\ \hat{z}=(0,0,1)$ and
the $SU(2)$ gauge
index $a$ runs 1, 2, 3.
$\eta_\alpha$, $W_\alpha$ and $\phi_\alpha$
($\alpha = 1,2$) are functions of
$(\rho, z)$.
Substituting \eqref{mansatz} into the deformed BPS equation \eqref{monopole},
we obtain 
\begin{eqnarray}
&&-\pa_3\eta_1+\eta_2W_2=\pa_\rho\phi_1- W_1\phi_2
+\epsilon\rho (\pa_\rho \eta_1+{\eta_1\over\rho}- W_1\eta_2),
\label{dbps1}
\\
&& -\pa_3\eta_2-\eta_1 W_2=
\pa_\rho\phi_2+ W_1\phi_1
+\epsilon\rho (\pa_{\rho}\eta_2+{\eta_2\over\rho}+
\eta_1 W_1),
\label{dbps2}
\\
&&
\pa_\rho\eta_1+{\eta_1\over\rho}- W_1\eta_2
=\pa_3\phi_1- W_2\phi_2
+\epsilon \rho (\pa_3\eta_1- \eta_2 W_2),
\label{dbps3}
\\
&&
\pa_\rho\eta_2+{\eta_2\over\rho}+\eta_1 W_1
=\pa_3\phi_2+ W_2\phi_1
+\epsilon\rho (\pa_3\eta_2+\eta_1 W_2),
\label{dbps4}
\\
&&
\pa_\rho W_2-\pa_3 W_1=-\eta_1\phi_2+\eta_2\phi_1,
\label{dbps5}
\end{eqnarray}
where $\del_\rho=\frac{\del}{\del\rho},\
\del_3 = \frac{\del}{\del z}$.
We have chosen the minus
sign in the BPS equation \eqref{monopole}.
These equations are invariant under
the gauge transformations \cite{Forgacs:1983sa}:
\begin{eqnarray}
W'_1=W_1+\pa_\rho \Lambda,
&\quad&
W'_2=W_2+\pa_3 \Lambda, \nonumber \\
\phi'_1=\cos\Lambda \phi_1+\sin\Lambda \phi_2,
&\quad&
\phi'_2=\cos\Lambda \phi_2-\sin\Lambda \phi_1,\\
\eta'_1=\cos\Lambda \eta_1+\sin\Lambda \eta_2,
&\quad&
\eta'_2=\cos\Lambda \eta_2-\sin\Lambda \eta_1,\nonumber
\end{eqnarray}
where $\Lambda$ is a function of $(\rho,z)$.
For $\epsilon=0$, these equations were
shown to be equivalent to the Ernst equation
and solved by Forg\'acs et al. \cite{Forgacs:1980yv}
by using the B\"acklund transformation technique.
We modify their gauge-fixing conditions to
\begin{align}
W_1=\eta_1,\quad W_2 = - \phi_1 -  \epsilon \rho \eta_1.
\end{align}
Then the equations \eqref{dbps1}
and \eqref{dbps5}
become equivalent. The equation
\eqref{dbps2} becomes
\begin{align}
-\del_3 \eta_2=
\del_{\rho}(\phi_2+\epsilon\rho\eta_2).
\end{align}
This can be solved by the following ansatz:
\begin{align}
\eta_2 = \frac{\del_\rho f}{f},\quad
\phi_2+\epsilon\rho\eta_2 = -\frac{\del_3f}{f},
\label{sol1}
\end{align}
where $f$ is a function of $(\rho,z)$
to be determined by other equations.
Under this ansatz, \eqref{dbps1} and
\eqref{dbps5} reduce to
\begin{align}
\del_\rho(f\phi_1+\epsilon\rho f\eta_1)
=-\del_3(f\eta_1).
\end{align}
This also can be solved as
\begin{align}
\eta_1 = \frac{\del_\rho\psi}{f},\quad
\phi_1+\epsilon\rho\eta_1=\frac{\del_3\psi}{f},
\label{sol2}
\end{align}
where $\psi$ is
also a function of $(\rho, z)$.
The remaining equations \eqref{dbps3} and \eqref{dbps4}
give the equations for $f$ and $\psi$:
\begin{align}
&f\left(\del_\rho^2 f + \frac{1}{\rho}\del_\rho f + \del_3^2 f\right)
-\left(\del_\rho f\right)^2-\left(\del_3 f\right)^2
+\left(\del_\rho \psi\right)^2+\left(\del_3 \psi\right)^2=0,
\label{ernst1}\\
&f\left(\del_\rho^2\psi + \frac{1}{\rho}\del_\rho\psi+\del_3^2\psi\right)
-2\del_\rho f\del_\rho\psi -2\del_3f\del_3\psi=0.
\label{ernst2}
\end{align}
They do not include the parameter
$\epsilon$ and coincide with
the Ernst equation \cite{Ernst:1967wx}.
Therefore we can construct
the deformed BPS monopole solution
from the potentials $f$ and $\psi$.

\begin{figure}[tbp]
\begin{center}
\begin{tabular}{cc}
\resizebox{80mm}{!}{\includegraphics{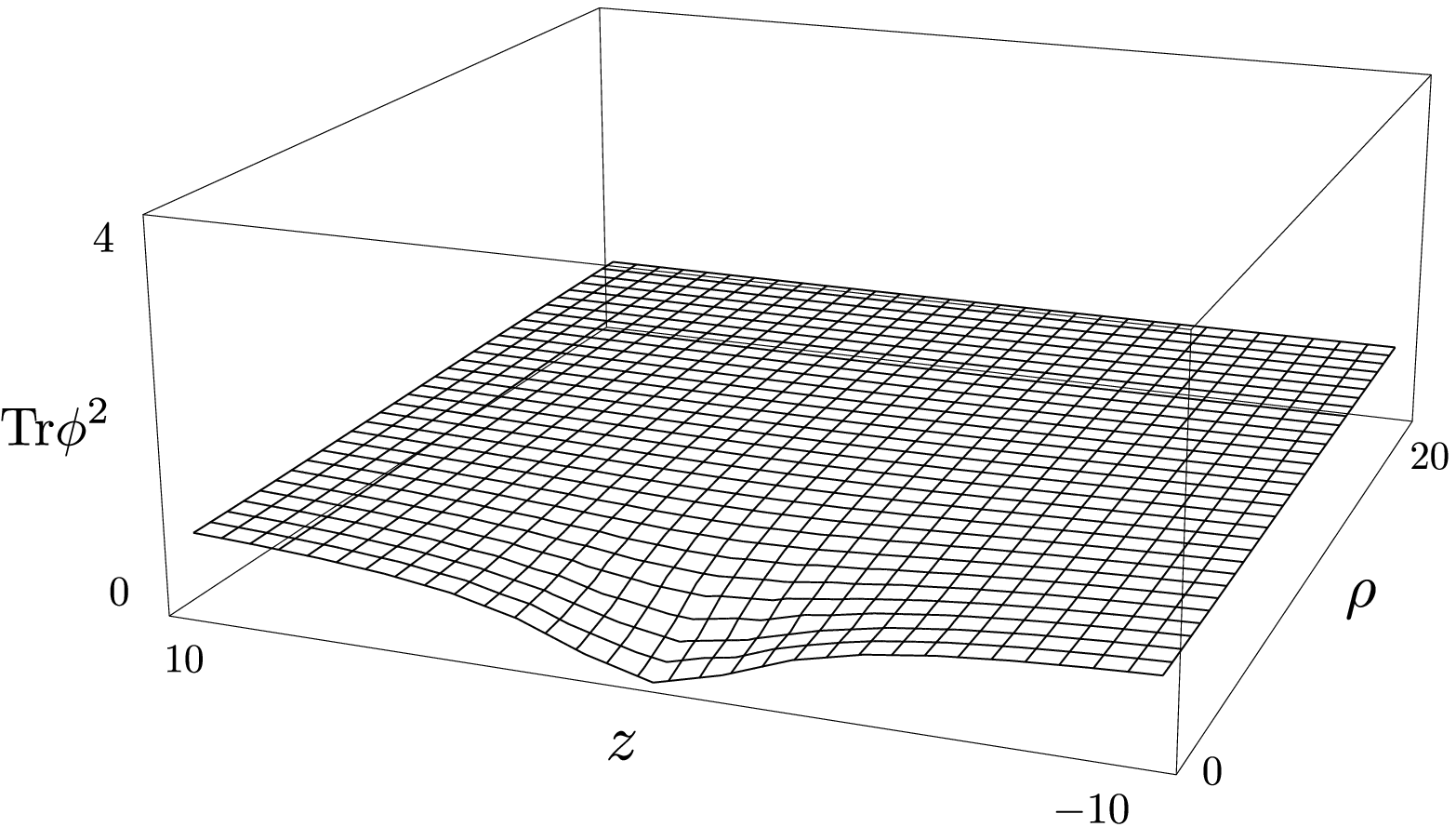}}&
\resizebox{80mm}{!}{\includegraphics{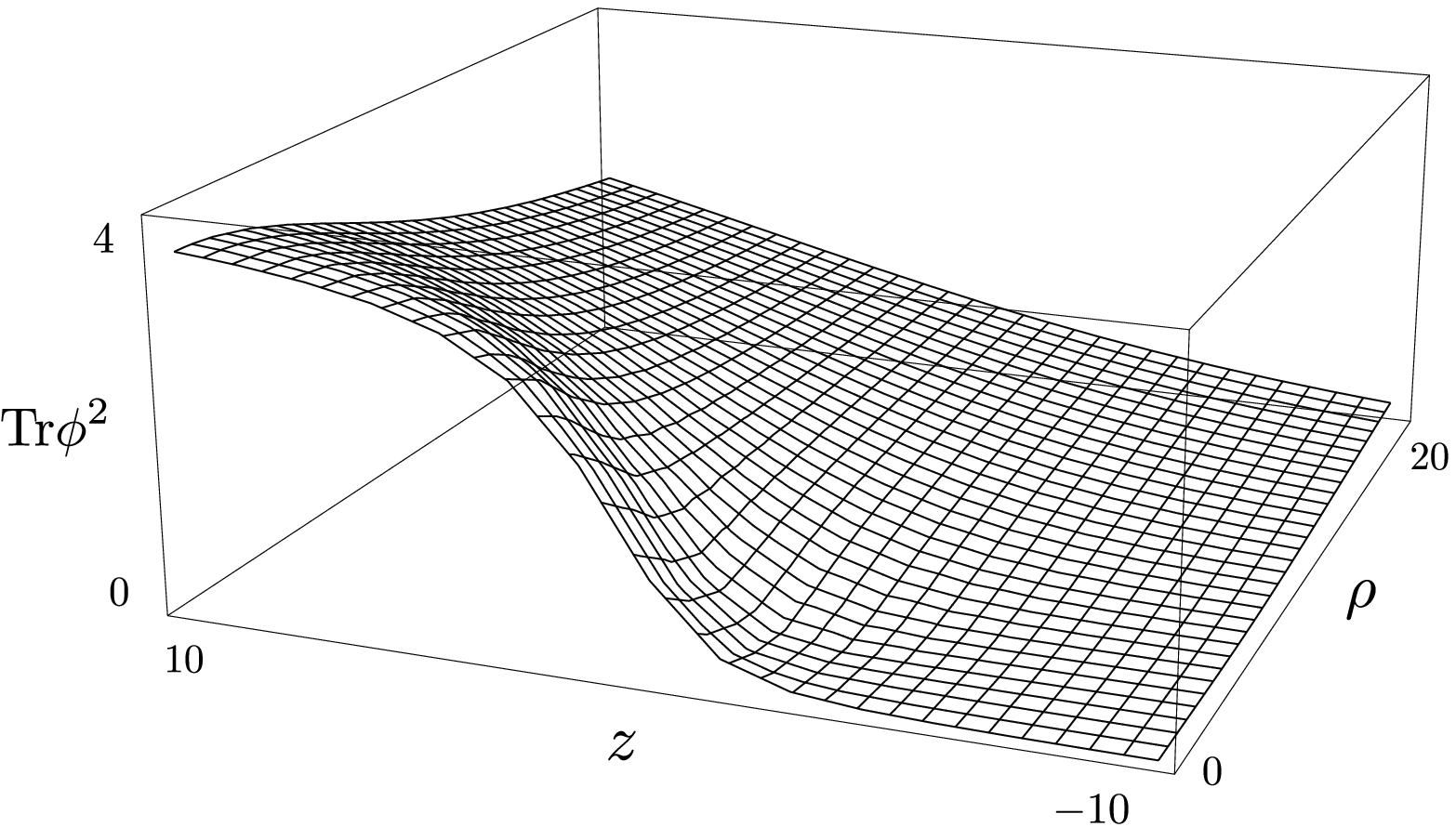}}\\
(a): $\epsilon=0$&(b): $\epsilon=1$
\end{tabular}
\end{center}
\caption{Profile of $\mathrm{Tr}\phi^2$ for $v=1$.}
\label{fig}
\end{figure}

In summary, we have obtained the following solution of the
deformed BPS monopole equation:
\begin{align}
\begin{aligned}
\eta_1=\frac{\del_\rho \psi}{f},
\quad
\eta_2=\frac{\del_\rho f}{f},&
\quad
W_1=\frac{\del_\rho \psi}{f},
\quad
W_2=\frac{\del_3 \psi}{f},
\\
\phi_1=-\frac{\del_3 \psi +  \epsilon \rho \del_\rho \psi}{f},&
\quad
\phi_2=-\frac{\del_3 f + \epsilon \rho \del_\rho f}{f},
\end{aligned}
\label{forg}
\end{align}
where $f$ and $\psi$ are the solutions
of the Ernst equations \eqref{ernst1}, \eqref{ernst2}.
We note that the gauge field in this solution is not
deformed by $\epsilon$. Its magnetic charge
can be rewritten without using the scalar field:
\begin{align}
Q_\mathrm{m}&=\int d^3 x B_i^a B_i^a.
\end{align}
Therefore the magnetic charge
is not deformed by $\epsilon$.
To demonstrate this fact explicitly, we construct
deformed one-monopole solution and evaluate
its magnetic charge. The deformed solution can be
constructed from the Ernst potential
for the undeformed one-monopole solution
\cite{Lohe:1977up,Forgacs:1980yv}
\begin{align}
\begin{aligned}
f_{1 \MP}=\rho /G,&\quad \psi_{1 \MP}=P/G, \\
G=\frac{r}{\sinh vr}+r\cosh vz\coth vr-z\sinh vz,&\quad
P=z\cosh vz-r\sinh vz \coth vr,
\end{aligned}
\label{1mp}
\end{align}
where $r=\sqrt{\rho^2+z^2}$ and $v=\langle\phi \rangle$
is the vacuum expectation value of the scalar field
in the undeformed theory with
spherical symmetry. In the deformed theory,
the asymptotic behavior of the solution depends
on the direction of the point at infinity.
Here we fix the value of $v$ as that of undeformed one.
Substituting \eqref{1mp} into \eqref{forg},
we obtain the deformed one-monopole solution.
We plot the gauge invariant quantity $\mathrm{Tr}\phi^2$
for $\epsilon=0$ and $\epsilon=1$ in Figure \ref{fig}. 
We see that for $\epsilon=0$, it is spherical symmetric and thus
isotropic in the $\rho$- and the $z$-direction.
For $\epsilon=1$, it receives the deformation
which breaks the spherical symmetry.
It increases in the region $z>0$
and thus becomes anisotropic.
As we mentioned above, its magnetic charge
remains undeformed:
\begin{align}
Q_\mathrm{m}&=4\pi v.
\label{defcc}
\end{align}

One can also solve the deformed monopole equation
in a different approach.
For $\epsilon=0$, we get the solution
of the form \cite{Prasad:1975kr,Bog}
\begin{align}
\begin{aligned}
\phi_1^{(0)}=\frac{v\rho}{r}H(r),
&\quad
\phi_2^{(0)}=\frac{vz}{r}H(r),\\
\eta_1^{(0)}=-\frac{z}{r^2}F(r),
&\quad
\tilde{\eta}_2^{(0)}=\frac{\rho}{r^2}F(r),\\
W_1^{(0)}=\frac{z}{r^2}F(r),
&\quad
W_2^{(0)}=-\frac{\rho}{r^2}F(r),
\label{bpssol}
\end{aligned}
\end{align}
where $\tilde{\eta}_2 = \eta_2 + 1/\rho $, and $F$, $H$
are defined by
\begin{align}
F(r) = 1-\frac{vr}{\sinh vr},\quad
H(r) = \coth vr -\frac{1}{vr}.
\end{align}
The superscript $(0)$ stands for the undeformed solution.
This is gauge-equivalent to the solution associated with \eqref{1mp}.
The deformed solution
can be obtained perturbatively by introducing
$\epsilon$-correction to this undeformed solution
as in the case of monopoles in non-commutative field theories
\cite{Hashimoto:1999az,Goto:2000zj}.
Thus we expand the deformed solution as
\begin{align}
X_\alpha = X^{(0)}_\alpha + \epsilon X^{(1)}_\alpha + \cdots ,
\end{align}
where $X_\alpha = \eta_1,\tilde{\eta_2},W_\alpha,\phi_\alpha$ and $\alpha = 1,2$. 
We assume that the $\epsilon$-correction has the following form:
\begin{align}
X^{(1)}_\alpha = \frac{1}{r^{n_1}}
P_{n_2}(\rho,z)
P_{n_3}(H(r),F(r)),
\end{align}
where we denote $P_{n}(s,t)$ by a polynomial
of $s$ and $t$ which has degree $n$.
When $n_1=3,\ n_2=3,\ n_3=1$ and we impose
the regularity at the origin and 
finiteness at the infinity on the solution,
we find that the corrections
\begin{align}
\phi^{(1)}_1=\frac{z\rho F}{r^2},\quad&
\phi^{(1)}_2=1-\frac{\rho^2F}{r^2},\\
\eta^{(1)}_1=\tilde{\eta}^{(1)}_2=&W^{(1)}_1=W^{(1)}_2=0,
\end{align}
satisfy the deformed BPS equation and there are no
higher-order corrections.
This solution has the same profile as in Figure \ref{fig}.
%We see that also in this solution 
In this approach 
the gauge field is
not deformed by $\epsilon$.
%, hence the charge remains undeforme
The charge remains undeformed
and coincides with \eqref{defcc}.

We now discuss the central charge in the SW theory.
In \cite{Mironov:2009uv, Mironov:2009dv},
the deformed prepotential has been obtained by the
deformation of the SW theory, which implies that the
central charge is expressed by the period integrals
on the deformed SW curve. The relation becomes
\begin{align}
Z = n_e a + n_m a_D,\quad a_D = \frac{\del\mathcal{F}}{\del a},
\label{zandf}
\end{align}
where $n_e$ and $n_m$ are the electric charge number
and the magnetic charge number respectively and $\mathcal{F}$ is
the deformed prepotential and $a = v/\sqrt{2}$. 
We have shown that the magnetic charge of the
BPS monopole is not deformed by $\epsilon$.
One can consider the BPS state with purely electric charge: the
W-bosons. By examining their mass under the condition
$\varphi = - \bar{\varphi}$ and $\Omega=-\bar{\Omega}$,
we see that the mass is not deformed by $\epsilon$.
Therefore, in the Manton ansatz and the perturbative approaches,
% for the BPS state with purely magnetic or electric charge
the central charge 
for the purely magnetic or electric BPS state
is not deformed by $\epsilon$.
It is not clear whether there are $\epsilon$-corrections for the
dyon state because the central charge formula (\ref{eq:central2})
contains the $\epsilon$-dependent term.
It is important to study the BPS dyon solution and its central charge in
order to determine the classical prepotential.
It is also an interesting problem to investigate the perturbative
corrections to the BPS mass since the perturbative part of the prepotential
receives the $\epsilon$-corrections \cite{Mironov:2009uv}.

In this paper we have studied the BPS monopole equations
using the approach of Forg\'acs et al.
It is interesting to study
the Nahm construction \cite{Nahm:1979yw,Bak:1999id} for monopoles
in the $\Omega$-background and its relation to the present approaches.
In the string theory, the undeformed Nahm construction
is naturally understood by the brane
configuration \cite{Diaconescu:1996rk}
and the $\Omega$-background is realized as a certain
$\mathcal{N}=2$ supergravity background
\cite{Billo:2006jm,Ito:2010vx}.
Hence if we find the deformed Nahm construction,
we may obtain the insight of the
stringy realization of the
monopole in the $\Omega$-background \cite{Hellerman:2011mv,Reffert:2011dp}.
The deformed monopole solutions that we have derived
would be helpful to find its construction.

\subsection*{Acknowledgment}
We are grateful to Y. Imamura for valuable discussions.
The work of K. I. is supported in part by Grant-in-Aid
for Scientific Research from the Japan Ministry of Education, Culture,
Sports, Science and Technology.
The work of S. S. is supported by the Japan Society
for the Promotion of Science (JSPS) Research Fellowship.

\end{document}